%% file: ITA_final.tex
\documentclass[10pt,conference]{IEEEtran}
\usepackage{amsmath,amsfonts, amssymb, mathrsfs}
\usepackage[dvips]{graphicx}
\usepackage[dvips]{color}
\def\SNR{\text{SNR}}

\def\EE{\mathbb E}
\def\eps{\varepsilon}

\def\PP{\mathbb P}
\def\RR{\mathbb R}
\def\Tr{\text{tr}}
\def\bw{\boldsymbol{w}}
\def\bx{\boldsymbol{x}}

\def\CM{C_{MIMO}}
\newtheorem{lem}{Lemma}[section]
\newtheorem{thm}{Theorem}[section]

\title{Linear Capacity Scaling in Wireless Networks: Beyond Physical Limits?}
\author{\authorblockN{Ayfer {\"O}zg{\"u}r, Olivier L\'ev\^eque}
\authorblockA{EPFL, Switzerland\\
              \{ayfer.ozgur, olivier.leveque\}@epfl.ch}
\and
\authorblockN{David Tse}
\authorblockA{
University of California at Berkeley\\
dtse@eecs.berkeley.edu}}
\begin{document}

\maketitle
\begin{abstract}
We investigate the role of cooperation in wireless networks subject to a spatial degrees of freedom limitation. To address the worst case scenario, we consider a free-space line-of-sight type environment with no scattering and no fading. We identify three qualitatively different operating regimes that are determined by how the area of the network $A$, normalized with respect to the wavelength $\lambda$, compares to the number of users $n$. In networks with $\sqrt{A}/\lambda\leq \sqrt{n}$, the limitation in spatial degrees of freedom does not allow to achieve a capacity scaling better than $\sqrt{n}$ and this performance can be readily achieved by multi-hopping. This result has been recently shown in \cite{FMM09}. However, for networks with $\sqrt{A}/\lambda> \sqrt{n}$, the number of  available degrees of freedom is $\min(n, \sqrt{A}/\lambda)$, larger that what can be achieved by multi-hopping. We show that the optimal capacity scaling  in this regime is achieved by hierarchical cooperation. In particular, in networks with $\sqrt{A}/\lambda> n$, hierarchical cooperation can achieve linear scaling.%, even in the considered deterministic line-of-sight type environments.
\end{abstract}

\section{Introduction}
Multi-hop is the communication architecture of current wireless
networks such as mesh or ad hoc networks. Packets are sent from each
source to its destination via multiple relay nodes. Each relay
decodes the packets sent from the previous relay and forwards them
to the next relay. Can more sophisticated cooperation between nodes
significantly increase capacity of such networks? This is an
important question concerning future communication architectures for
such networks, and information theory has been  brought to bear to
try to shed some light on this question. Adopting the scaling law
formulation of Gupta and Kumar \cite{GK00}, much focus has been on
the asymptotic regime where the number of nodes is large. Two
diametrically opposite answers have emerged:

\begin{itemize}
\item 1) Capacity can be significantly improved when nodes form distributed MIMO arrays via an intelligent cooperation architecture
\cite{AS06,OLT07}. The total degrees of freedom in the network is
$n$, the number of nodes, and in regimes where power is not a
limiting factor, the capacity can scale almost linearly with $n$.

\item 2) The total degrees of freedom in the network is not $n$ but
is actually upper bounded by $\sqrt{n}$ due to the spatial
constraints imposed by the physical channel \cite{FMM09}.
Nearest-neighbor multi-hop is optimal to achieve this scaling
\cite{GK00}.

\end{itemize}

This is no mathematical contradiction between these two sets of
results. They are based on two different channel models. The key
difference is the assumption on the phases of the channel gains
between the nodes. \cite{AS06,OLT07} assume that the phases are
uniform and independent across the  different channel gains.
\cite{FMM09}, on the other hand, starts from physical principles and
regards the phases as functions of the locations of the nodes. While
the physical channel model used in \cite{FMM09} is more fundamental,
the i.i.d. phase model is also widely accepted in wireless
communication engineering, particularly for nodes in far field from
each other. Is there a way to reconcile the two sets of results?

 A deeper look at \cite{FMM09} provides a clue. The spatial degrees
of freedom limitation in \cite{FMM09} is actually dictated by the
{\em diameter} of the network rather than the number of nodes. More
precisely, the spatial degrees of freedom in the network are limited
by $\sqrt{A}/\lambda$, where $A$ is the area of the network and
$\lambda$ is the carrier frequency. This number can be heuristically
thought of as an upper bound to the total degrees of freedom in the
network as a whole and puts a limitation on the maximum possible
cooperation gain. The conclusion that the capacity scales like
$\sqrt{n}$  comes from the assumption that the {\em density} of
nodes is fixed as the number of nodes $n$ grows, so that
$\sqrt{A}/\lambda$ is proportional to $\sqrt{n}$. But for actual
networks, there can be a huge  difference between $\sqrt{A}/\lambda$
and $\sqrt{n}$. Take an example of a network serving $n=10,000$
users on a campus of $1$ km$^2$, operating at $3$ GHz:
$\sqrt{A}/\lambda=10000$, while $\sqrt{n}$ is only $100$, two orders
of magnitude smaller. So while multi-hop can achieve a total
throughput of the order of $100$ bits/s/Hz, there is still a lot of
potential for cooperation gain, since the spatial degrees of freedom
upper bound is $10,000$.

So the ultimate cooperation gain is limited by $\sqrt{A}/\lambda$,
while multi-hop performance depends on the number of nodes $n$ only
and not on $\sqrt{A}/\lambda$. But the number of nodes and the area
are two independent parameters of a network, each of which can take
on a wide range of values. To yield a complete picture of whether
cooperation can help, the key is to remove the artificial coupling
between these two parameters and analyze the capacity in terms of
the two parameters {\em separately}. This is the goal of the present
paper. We focus on a physical channel model similar to that used in
\cite{FMM09}, but with only a line-of-sight channel between each
pair of nodes, a case in which spatial limitation is expected to be
the most severe. Our main result is that in the regime when $n$ and
$A/\lambda$ are both large, the capacity of the network is
approximately
\begin{equation}
\max\left ( \sqrt{n}, \min ( n, \frac{\sqrt{A}}{\lambda} )\right).
\label{eq:main_result}
\end{equation}

Accordingly, the optimal operation of the network falls into three different operating regimes:
\begin{itemize}
\item 1) $ \sqrt{A}/\lambda \le \sqrt{n}$: The number of spatial degrees of freedom is too small, cooperation is useless and nearest neighbor
multi-hopping is  optimal.
\item 2) $\sqrt{A}/\lambda > n$: The number of spatial degrees of freedom is $n$, cooperation is very useful, and the optimal performance can be
achieved by the same hierarchical cooperation scheme introduced in
\cite{OLT07}. Spatial degree of freedom limitation does not come
into play and the performance is {\em as though} the phases are
i.i.d. uniform across the nodes.
\item 3) $ \sqrt{n} \le \sqrt{A}/\lambda \le n$: The number of degrees of
freedom is smaller than $n$, so the spatial limitation is felt,  but
larger than what can be achieved by simple multi-hopping. A
modification of the hierarchical cooperation scheme achieves optimal
scaling in this regime.

\end{itemize}

Regime (1) is essentially the conclusion of \cite{FMM09}; regime (2)
is essentially the conclusion of \cite{OLT07} (in the case when
power is not a limiting factor). Thus, the validity of the results
in these papers is not universal but depends on the relationship
between $n$ and $A/\lambda$. The {\em upper bound} of
$\sqrt{A}/\lambda$ on the spatial degrees of freedom of the network
is already established by \cite{FMM09}. The main technical
contributions of the present paper are two-folded: 1) we show that
there are actually $\min(n, \sqrt{A}/\lambda)$ spatial degrees of
freedom available in the physical channel model when
$\sqrt{A}/\lambda \ge \sqrt{n}$; 2) we show that hierarchical
cooperation can achieve these degrees of freedom.

Both mathematically and philosophically, the present paper follows
the same spirit of \cite{OJTL10}. \cite{OJTL10} advocates a shift of
the ``large networks'' research agenda from seeking a single
``universal'' scaling law, where the number of nodes $n$ scales with
all systems parameters coupled with $n$ in a specific way, to
seeking a {\em multi-parameter family} of scaling laws, where the
key parameters are decoupled and many different limits with respect
to these parameters are taken. A single scaling law with a
particular coupling between parameters is often arbitrary and too
restrictive to cover the wide ranges that the multiple parameters of
the network can take on. The specific parameters that were decoupled
in \cite{OJTL10} were the number of nodes and the amount of power
available. The current paper follows the approach of \cite{OJTL10},
but focuses on the number of nodes and the area of the network,
while assuming there is a sufficient amount of power available that
it is not limiting performance. A future goal of this research
program is to investigate the dependence of the capacity on the number of nodes, the area of the network and the amount of power all together.

\section{Model}
There are $n$ nodes with transmitting and receiving capabilities that are uniformly and independently distributed in a rectangle of area $\sqrt{A}\times\sqrt{A}$. Each node has an average transmit power budget of $P$ Watts and the network is allocated a total bandwidth of $W$ Hertz around a carrier frequency of $f$, $f\gg W$. Every node is both a source and a destination for some traffic request. The sources and destinations are randomly paired up one-to-one into $n$ source-destination pairs without any consideration on node locations. Each source wants to communicate to its destination at the same rate $R$ bits/s/Hz. The aggregate throughput of the system is $T = n R$.

We assume that communication takes place in free-space line of sight type environment and the complex baseband-equivalent channel gain between  node $i$ and node $k$ is given by
\begin{equation}
\label{eq:ch_model0} 
H_{ik} = \sqrt{G}\,\,\frac{e^{j2\pi r_{ik}/\lambda}}{r_{ik}}
\end{equation}
where $r_{ik}$ is the distance between the nodes $i$ and $k$  and $\lambda$ is the carrier wavelength. Note that the locations of the users are drawn randomly but remain fixed over the duration of the communication. Therefore for a given realization of the network, the channel coefficients in (\ref{eq:ch_model0}) are deterministic. 

The parameter $G$ is given by the Friis' formula, 
\begin{equation}\label{G}
 G = \frac{G_{Tx} \cdot G_{Rx} \cdot \lambda^2}{ 16 \pi^2}, 
\end{equation}
where $G_{Tx}$ and $G_{Rx}$ are the transmitter and receiver antenna
gains respectively. The discrete-time complex baseband signal received by node $i$ at time $m$ is given by
\begin{equation}
\label{eq:ch_model2} 
Y_i[m]=\sum_{\substack{k=1,\,k\neq i}}^{n}H_{ik}X_k[m]+Z_i[m]
\end{equation}
where $X_k[m]$ is the signal sent by node $k$ at time $m$ subject to an average power constraint 
$$
\EE(|X_k|^2)\leq P/W
$$
and $Z_i[m]$ is complex white circularly symmetric Gaussian noise of variance $N_0$. The model in (\ref{eq:ch_model0}), \eqref{G}   corresponds to free-space propagation. It is equivalent to the model in Section IV of \cite{FMM09} but with no scatterers. We consider the case of no scatterers since the spatial degrees of freedom limitation is expected to be most severe in this case.% Note that the model in  \eqref{eq:ch_model} is dramatically different from the i.i.d fading channel model of \cite{OLT07}. Indeed, here there is no fading, the channel coefficients are determined deterministically by the node positions.
%\footnote{Through private communication, we have been informed  of an independent work \cite{LC10} that similarly investigates the performance of the hierarchical cooperation scheme under this LOS channel model.}

It has been shown in \cite{OJTL10} that a wireless adhoc network is power-limited when the long-range $\SNR$ in the network is smaller than $0$ dB and the long range $\SNR$ has been identified as 
\begin{equation}\label{def:SNRl}
\SNR_l:= n\frac{GP}{N_0\,W\,(\sqrt{A})^{\alpha}}.
\end{equation}
For the current case $\alpha=2$, which implies that $\SNR_l=\SNR_s$, where $\SNR_s$ is the $\SNR$ in a point-to-point transmission over the typical nearest neighbor distance in the network. (See also \cite{Oz09}.) In the present paper, our goal is to concentrate on the effect of the spatial degrees of freedom limitation on the capacity of wireless adhoc networks. To be able to solely concentrate  on this factor, we assume there is no power limitation in our network. Formally, we assume that $P$ and $W$ are such that
\begin{equation}\label{eq:SNR_l}
\SNR_l>\,0\,\,\text{dB}, 
\end{equation}
for every $A$ and $n$. For the current case of $\alpha=2$, the condition can be equivalently stated as $\SNR_s>0$ dB. When this condition fails to hold, the network becomes power limited and the behavior of the capacity as well as optimal operation can be significantly different. 

\section{Main result}

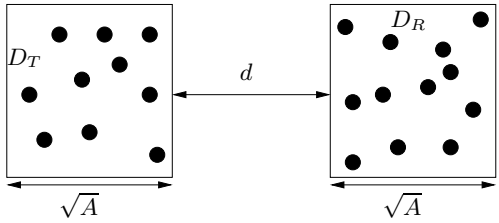
\begin{figure}[tbp]
\begin{center}
\input{squares.pstex_t}
\end{center}
\caption{Two square clusters of area $A_c$ separated by a distance $d$.}
\label{squares}
\end{figure}

The main result of \cite{FMM09} is to show that under the network and the channel model described in the previous section with the additional restriction $A=n$, the capacity of the wireless network is upper bounded by
$$
T \le K_1 \, \sqrt{n}\, (\log n)^2,
$$
with high probability\footnote{With probability $1$ as $n\rightarrow\infty$.}, where $K_1>0$ is a constant independent of $n$. Coupling the area of the network a priori with the number of nodes in the network is restrictive and does not allow to deduce the nature of the limitation imposed here. A relatively straightforward generalization of the analysis in \cite{FMM09} gives the following result. Let us define the normalized area of the network with respect to the wavelength $\lambda$ as, 
$$
A_0:=\frac{A}{\lambda^2}.
$$
Under the network and channel model described in the previous section, the capacity of the wireless network is upper bounded by
$$
T \le \left\{\begin{array}{ll}K_1 \, \min\left(n\,(\log n)^2\,,\,\sqrt{A_0}\,\left(\log \sqrt{A_0}\right)^2\right) & \text{if }A_0 > n\\
K_1 \, \sqrt{n}\,(\log n)^2 & \text{if } A_0 \le n \end{array} \right.
$$
with high probability %\footnote{With probability $1$ as $n, A\rightarrow\infty$.} 
where $K_1>0$ is a constant independent of $n$ and $A_0$. For $A_0 \le n$, this result says that the maximum achievable capacity is of order $\sqrt{n}$, which is achievable by a simple multi-hopping scheme \cite{GK00}. For $A_0>n$, the achievability remains an open issue so far. 
\smallbreak

The following theorem is the main contribution of the present paper.
\begin{thm} \label{sqrtA0}
Consider the network and the channel model described in the previous section and assume $A_0>n$, the total throughput achieved by hierarchical cooperation is lower bounded by,
$$
T \ge K_2 \, \left(\min(n, \sqrt{A_0})\right)^{1-\varepsilon}
$$
with high probability, for any $\varepsilon>0$ and a constant $K_2>0$ independent of $n$ and $A_0$.
\end{thm}

The theorem can be interpreted as follows: When $A_0>n^2$, hierarchical cooperation can achieve an aggregate throughput $T>K_2\,n^{1-\varepsilon}$ for any $\varepsilon>0$. When $A_0<n^2$, hierarchical cooperation can achieve an aggregate throughput $T>K_2\,A_0^{1/2+\varepsilon}$. Note that this throughput is larger than $\sqrt{n}$, when  $A_0>n$. 

\section{Hierarchical Cooperation in LOS environments}
The proof of Theorem \ref{sqrtA0} relies on the following lemma which establishes a lower bound on the capacity of a MIMO transmission between two clusters of nodes. For notational simplicity, in the sequel we assume that all the distances in the network are normalized with respect to the carrier wavelength $\lambda$. Note that when the distances $r_{ik}$ are expressed in wavelength units, the channel model in \eqref{eq:ch_model0}, \eqref{G} takes the simplified form,
\begin{equation}\label{eq:ch_model} 
H_{ik} = \sqrt{G}\,\,\frac{e^{j2\pi r_{ik}}}{r_{ik}}
, \qquad  G = \frac{G_{Tx} \cdot G_{Rx}}{ 16 \pi^2}.
\end{equation} 

\begin{lem} \label{Aoverd}
Consider two square clusters of area $A_c$ separated by a distance $d$ (see Figure~\ref{squares}), with each cluster containing $M$ nodes distributed uniformly at random over $A_c$. Let $\sqrt{A_c} \leq d \leq A_c$,  and the nodes in the transmit cluster $D_T$ perform independent signalling with power $P_0/M$ such that 
\begin{equation}\label{powerMIMO}
\frac{GP_0}{N_0 W d^2}\,>\, 0\, \text{dB}.
\end{equation}
Then the capacity of the MIMO channel from the transmit cluster $D_T$ to the receive cluster $D_R$ is lower bounded by
\begin{align*}
\CM &=\left( \log\det\left(I+ \frac{P_0}{N_0 W }\,\frac{1}{M}\, HH^\dagger\right) \right)\\&\ge K_3 \, \min \left(M,\frac{A_c/d}{\log(A_c/d)} \right)
\end{align*}
with high probability%\footnote{With probability $1$ as $M\rightarrow\infty$.} 
for some constant $K_3>0$ independent of $M$, $A_c$ and $d$.
\end{lem}

The lemma is the analog of Lemma~4.3 in \cite{OLT07} which lower bounds the capacity of a MIMO transmission between two clusters of nodes under the i.i.d. phase model. With i.i.d. phases, the capacity of the MIMO transmission scales linearly in $M$. The condition \eqref{powerMIMO} ensures that the MIMO transmission is not power limited. For the LOS case, we have the additional term $\frac{A_c/d}{\log(A_c/d)}$, which corresponds to the spatial degrees of freedom between the two clusters. When this term is smaller than $M$, the capacity of the MIMO transmission is not any more linear in $M$. This in turn degrades the performance of the hierarchical cooperation scheme which is based on such MIMO transmissions. 

The capacity of a MIMO transmission between two clusters under the current
LOS channel model has been investigated earlier in \cite{LC08}. The result stated
in Theorem~1 of \cite{LC08} is equivalent to Lemma~\ref{Aoverd} above. However, the proof of Theorem~1 in \cite{LC08} is based on an approximation which is not fully justified. Through private communication, we have been informed of a
follow-up work \cite{LC10} by the same authors, that similarly to our current
paper investigates the performance of the hierarchical cooperation scheme
under the LOS channel model.

Next, we investigate the performance of the hierarchical cooperation scheme and show how Lemma~\ref{Aoverd} allows to prove the result in Theorem~\ref{sqrtA0}. The core of the proof is the following recursion lemma.

\begin{lem}\label{lem:rec}
Consider a network of $n$ nodes uniformly distributed over an area $A_0>n$  and the available power $P$ per node satisfies (\ref{eq:SNR_l}). Assume that there exists a communication scheme for this network that achieves an aggregate throughput
$$
T\geq K_4\, \min(n,\sqrt{A_0})^{b}
$$
with high probability for some $0\leq b <1$ and a constant $K_4>0$ independent of $n$ and $A_0$. 

Then, we can construct another scheme for this network  that  achieves a higher aggregate throughput
$$
T\geq K_5\, \min(n,\sqrt{A_0})^{\frac{1}{2-b}-\varepsilon_1}
$$
with high probability for any $\varepsilon_1>0$ and a constant $K_5>0$ independent of $n$ and $A_0$. 
\end{lem}

\smallbreak

As soon as we have a scheme to start with, Lemma~\ref{lem:rec} can be applied recursively, yielding a scheme that achieves higher throughput at each step of the recursion. Note that $\frac{1}{2-b}>b$ for $0\leq b<1$. We first show that a simple time-sharing strategy between the source-destination pairs (TDMA) satisfies the conditions of the lemma with $b=0$. Note that with TDMA, each source node transmits only a fraction $1/n$ of the total time of communication. Hence when active, each source node can transmit with elevated power $nP$ and still satisfy its average power constraint $P$. This yields an SNR larger than $\SNR_l$ in (\ref{def:SNRl}) for each transmission, hence a constant rate. Therefore, the aggregate throughput achieved by TDMA is constant independent of $n$ and $A_0$. 

Starting with TDMA, $b=0$, and applying Lemma~\ref{lem:rec} recursively $h$ times, we get a hierarchical scheme that achieves an aggregate throughput of order $\min(n,\sqrt{A_0})^{\frac{h}{h+1}-\varepsilon_1^\prime}$ for any $\varepsilon_1^\prime>0$. Therefore given any $\varepsilon>0$, we can choose $\varepsilon_1^\prime=\varepsilon/2$ and $h$ such that $\frac{h}{h+1}\geq 1-\varepsilon/2$ and we a get a scheme that achieves the performance in Theorem~\ref{sqrtA0}.\hfill$\square$
 
\medbreak 
  
\textit{Proof of Lemma~\ref{lem:rec}:} We will prove the lemma by concentrating separately on the two cases $A_0>n^2$ and $n<A_0\leq n^2$. In the first case, we provide a brief overview of the three-phase scheme from Lemma~3.1 in \cite{OLT07} and verify that it achieves the same performance in \cite{OLT07} under the current deterministic phase model. The reader should refer to \cite{OLT07} for a precise analysis. For the case $n<A_0 \leq n^2$, a modification of the scheme is required to achieve the performance given in Lemma~\ref{lem:rec}.

\subsection{ $A_0>n^2$ } \label{HC1}

Let us divide the network into square clusters of area $A_c$. Each cluster contains approximately $M=\frac{A_c}{A_0}n$ nodes. A particular source node $s$ sends $M$ bits to its destination node $d$ in three steps:
\begin{itemize}
\item [(S1)] Node $s$ first distributes its $M$ bits among the $M$ nodes in its cluster, one bit for each node;
\item [(S2)] These nodes together can then form a distributed transmit antenna array, sending the $M$ bits {\em simultaneously} to the destination cluster where $d$ lies;
\item [(S3)] Each node in the destination cluster observes the MIMO transmission in the previous phase; it quantizes each observation to $Q$ bits, with a fixed $Q$, and ships them to $d$, which can then do joint MIMO processing of all the quantized observations and decode the $M$ transmitted bits from $s$.
\end{itemize}
From the network point of view, all source-destination pairs have to
eventually accomplish these three steps. Step 2 is long-range
communication and only one source-destination pair can operate at
a time. Steps 1 and 3 involve local communication and can be
parallelized across clusters. 

Since there are $M$ source nodes in every cluster, this gives a total traffic of exchanging $M(M-1)\sim M^2$ bits inside each cluster in phase 1. We can handle this traffic by setting up $M$ sub-phases, and assigning $M$ pairs in each sub-phase to communicate their $1$ bit. The traffic to be handled at each sub-phase is similar to our original network communication problem with $n$ users on an area $A_0$, but now instead, we have $M$  users on area $A_c$. We handle this traffic using the communication scheme given in Lemma~\ref{lem:rec}. Note that if this scheme achieves an aggregate throughput $K_4\,\min(n,\sqrt{A_0})^{b}$ in the network of $n$ nodes and area $A_0$, it will achieve an aggregate rate $K_4\,\min(M,\sqrt{A_c})^{b}$ inside the clusters of $M$ nodes and area $A_c$.\footnote{We ignore the performance loss due to  inter-cluster interference since it does not change the scaling law. The reader is referred to \cite{Oz09} for details.} This can be verified by checking that the clusters of $M$ nodes and area $A_c$ satisfy the conditions of the lemma. We have $A_c>M$ for the clusters if $A_0>n$ for the original network and 
$$
\SNR_l(M,A_c)=M\frac{GP}{N_0\,W\,A_c}=\SNR_l\,>\,0\,\text{dB}
$$
if $P$ satisfies (\ref{eq:SNR_l}). Moreover when $A_0>n^2$, we have $A_c>M^2$, so the performance of the scheme is $K_4\,\min(M,\sqrt{A_c})^{b}=M^b$. The traffic in the third phase is handled similarly to the first phase. Then, we need:
\begin{itemize}
\item $M^{2-b}/K_4$ time slots to complete phase 1 all over the network; We handle the traffic in $M$ subphases, each subphase is completed in $M^{1-b}/K_4$ time-slots.
\item $n/K_3$ time-slots to complete the successive MIMO transmissions in the second phase, \emph{if the distributed MIMO transmissions between any two clusters can achieve a rate of $K_3M$ bits/time-slot}; We perform one MIMO transmission for each of the $n$ source-destination pairs in the network.
\item $QM^{2-b}/K_3K_4$ time slots to complete phase 3 all over the network; The traffic in the third phase is symmetrical to the traffic in the first phase, but larger by a factor of $Q/K_3$. This factor  comes from the fact that each MIMO transmission lasts $1/K_3$ time slots, and each of the corresponding $1/K_3$ observations is quantized to $Q$ bits.
\end {itemize}

In \cite{OLT07}, it is shown that each destination node is able to decode the transmitted bits from its source node from the $M$ quantized signals
it gathers by the end of Phase 3. Thus, the aggregate throughput achieved by the scheme can be calculated as follows: each source node is able to transmit $M$ bits to
its destination node, hence $nM$ bits in total are delivered to
their destinations in $M^{2-b}/K_4+n/K_3+QM^{2-b}/K_3K_4$ time slots, yielding an
aggregate throughput of 
\vspace*{-0.1cm}
$$ \frac{nM}{M^{2-b}/K_4+n/K_3+QM^{2-b}/K_3K_4}\quad\text{bits/time-slot.}
$$ Choosing $M=n^{\frac{1}{2-b}}$ to maximize this expression yields an aggregate throughput $T=K_5n^{\frac{1}{2-b}}$ for a constant $K_5>0$. 

\smallbreak

Note that this throughput can only be achieved if the MIMO transmissions in phase 2 achieve a rate linear in $M$. The rate of the MIMO transmissions are lowerbounded in Theorem~\ref{Aoverd} for the deterministic phase model under certain conditions. The cluster areas and the separation between the clusters should satisfy the condition $\sqrt{A_c} \leq d \leq A_c$ and the users should transmit with power satisfying condition (\ref{powerMIMO}). It is easy to verify that $\sqrt{A_c} \leq d \leq A_c$. Note that $\sqrt{A_c} \leq d$ is always true unless the communicating clusters are neighbors.\footnote{The special case of neighboring clusters is excluded from the current discussion and can be handled separately as in \cite{OLT07}.} Let us verify that the power condition (\ref{powerMIMO}) for the MIMO transmission can be satisfied under the average power constraint $P$ per node satisfying (\ref{eq:SNR_l}). In the second phase, the MIMO transmissions between clusters are performed successively and each node in the network transmits only $M/n$ of the time. Therefore when active, each node can transmit with elevated power $nP/M$%, or power $P_0/M$ with $P_0=nP$, 
and still satisfy its average power constraint $P$. Observe that if $P_0=nP$, the condition (\ref{powerMIMO}) is satisfied given (\ref{eq:SNR_l}) and the  fact that $d<\sqrt{A_0}$.

%Let us verify that we also satisfy the condition $\sqrt{A_c} \leq d \leq A_c$ for our choice $M=n^{\frac{1}{2-b}}$.  On the other hand, $d<\sqrt{A_0}$ and if $M=n^{\frac{1}{2-b}}$, $A_c=A_0\,n^{-\frac{1-b}{2-b}}$. Therefore $d\leq A_c$ if $A_0>n$ which is the case here.

Therefore, Theorem~\ref{Aoverd} lowerbounds the rate of the MIMO transmissions in the second phase. The lower bound is linear in $M$ if $\frac{A_c/d}{\log(A_c/d)}\geq M$. If $A_0>n^2$, using $A_c=\frac{MA_0}{n}$ and $d\geq\sqrt{A_0}$, we obtain for sufficiently large $M$,
\begin{equation*}
\frac{A_c/d}{\log(A_c/d)}\geq \frac{M\sqrt{A_0}/n}{\log(M\sqrt{A_0}/n)}\geq M^{1-\varepsilon_1}
\end{equation*}
for any $\varepsilon_1>0$. The $\varepsilon_1$ is introduced to compensate for the logarithmic term and in turn yields an $n^{-\varepsilon_1}$ degradation in the overall throughput as stated in Lemma~\ref{lem:rec}. This concludes the proof of the lemma for networks with $A_0>n^2$.

\subsection{ $n<A_0\leq n^2$ } \label{HC2}

In the case $n<A_0\leq n^2$, the proof of the lemma differs from the earlier case  $A_0>n^2$  in two aspects. When $n<A_0\leq n^2$, the MIMO transmissions between the clusters are limited in spatial degrees of freedom. More precisely, in Theorem~\ref{Aoverd}, the performance is lower bounded by the second term $\frac{A_c/d}{\log(A_c/d)}$ and it is not anymore linear in $M$. This fact requires a modification in the operation of this phase.

The second difference is the following: We have seen that when $A_0>n^2$ for the original network, we have $A_c>M^2$ for the smaller clusters. In other words, when the network is not spatial degrees of freedom limited at the largest scale, it is not spatial degrees of freedom limited at any scale. In the current case, when $n<A_0\leq n^2$, the network is limited in spatial degrees of freedom at the largest scale, but the smaller clusters may or may not be spatial degrees of freedom limited. More precisely, for a cluster of smaller size, we can either have $M<A_c\leq M^2$ or $A_c>M^2$. This fact requires a more careful analysis. In particular, we separately consider the two cases $n<A_0\leq n^{\frac{2(4-b)}{5-2b}}$ and $n^{\frac{2(4-b)}{5-2b}}<A_0\leq n^2$.

\bigskip

\subsubsection{$n^{\frac{2(4-b)}{5-2b}}<A_0\leq n^2$}

As before, we divide the network into clusters of area $A_c$ that contain $M=n\,A_c/A_0$ nodes and the goal again is to accomplish steps S1-S2-S3 for every source-destination pair in the network. We choose the cluster size in the following particular way,
\begin{equation}\label{Ac_1}
M= n^{\frac{2}{2-b}}\,A_0^{-\frac{1}{2(2-b)}}.
\end{equation}
This is a valid choice in the sense that $M<n$, in particular $M<n^{\frac{1}{2-b}}$ given the condition $A_0\,\leq\, n^2$
for the network. The condition $n^{\frac{2(4-b)}{5-2b}}\,<A_0$ ensures that $A_c> M^2$. Therefore as before, the scheme given in the hypothesis of Lemma~\ref{lem:rec} achieves an aggregate throughput $K_4\,\min(M,\sqrt{A_c})^{b}=M^b$ when used inside the clusters of area $A_c$ and number of nodes $M$. We use this scheme to handle the traffic inside the clusters in phases 1 and 3 as before. In the second phase, the MIMO transmissions achieve a rate
$$
\frac{A_c/d}{\log(A_c/d)}\geq \frac{A_c/\sqrt{A_0}}{\log(A_c/\sqrt{A_0})}.
$$
This implies that in the second phase, the MIMO transmissions for each source-destination pair  can not be completed in constant number of time-slots as before. In order for these MIMO transmissions of lower rate not to result in too many MIMO observations in the third phase containing a small number of degrees of freedom, we introduce the following modification to step (S2). Let
\begin{equation}\label{Mprime}
M^\prime=\frac{A_c/\sqrt{A_0}}{\log(A_c/\sqrt{A_0})}.
\end{equation}
We randomly divide the $M$ nodes in the source cluster to $M/M^\prime$ groups each containing $M^\prime$ nodes. We do the same division also in the destination cluster. We randomly associate one-to-one the $M/M^\prime$ groups in the source cluster with the $M/M^\prime$ groups in the destination cluster. The earlier $M\times M$ MIMO transmission between the source and the destination cluster is now divided into $M/M^\prime$ successive MIMO transmissions, each of size $M^\prime\times M^\prime$. In each of these $M^\prime\times M^\prime$ MIMO transmissions, a group of $M^\prime$ nodes in the source cluster are simultaneously transmitting their bits to their corresponding group in the destination cluster. Note that these  
$M^\prime\times M^\prime$ MIMO transmissions are not limited in spatial degrees of freedom, precisely due to our choice for $M^\prime$ in (\ref{Mprime}). We will later verify that these $M^\prime\times M^\prime$ MIMO transmissions achieve a rate  $K_3\,M^\prime$. If this is the case, we need:
\begin{itemize}
\item $M^{2-b}/K_4$ time slots to complete phase 1 all over the network; %Recall that we handle the traffic in $M$ subphases, each subphase is completed in $M^{1-b}/K_4$ time-slots using the scheme in Lemma~\ref{lem:rec}.
\item $n\times M/M^\prime\times 1/K_3$ time-slots to complete the successive MIMO transmissions in the second phase, \emph{if the distributed $M^\prime\times M^\prime$ MIMO transmissions between any two groups can achieve a rate of $K_3M^\prime$ bits/time-slot}; %We have $n$ source-destination pairs, the MIMO transmission between each source destination pair is divided to $M/M^\prime$ succesive $M^\prime\times M^\prime$ MIMO transmissions, each of duration $1/K_3$ time-slots.
\item $QM^{2-b}/K_3K_4$ time slots to complete phase 3 all over the network; Note that although each cluster receives $M\times M/M^\prime$ MIMO transmissions in total, $M/M^\prime$ MIMO transmissions per each destination node in the cluster, each node has one MIMO observation of duration $1/K_3$ time-slots for each of the other nodes. The modification in the second phase is precisely made to ensure this fact.
\end {itemize}

Thus, the aggregate throughput achieved by the scheme is given by 
\vspace*{-0.3cm}
\begin{equation}\label{eq:aggthr}
\frac{nM}{M^{2-b}/K_4+nM/M^\prime K_3+QM^{2-b}/K_3K_4}
\end{equation}bits per time-slot. It can be verified that for the choice of the cluster size in \eqref{Ac_1}, we have
\vspace*{-0.1cm}
$$
M^{2-b}=\frac{nM}{A_c/\sqrt{A_0}}.
$$
The three terms in the denominator of \eqref{eq:aggthr} are order-wise equal or in other words, \eqref{Ac_1} is the cluster size that maximizes the throughput expression in \eqref{eq:aggthr}. This yields an aggregate throughput 
\vspace*{-0.3cm}$$T =K_5M^\prime=K_5 \frac{A_c}{\sqrt{A_0}} A_0^{-\varepsilon_1}=K_5\, n^{\frac{b}{2-b}}\,A_0^{\frac{1-b}{2(2-b)}}A_0^{-\varepsilon_1},$$  for a constant $K_5>0$ and for any $\varepsilon_1>0$, which is introduced to compensate for the logarithmic term in (\ref{Mprime}). It can be verified that when $A_0\leq n^2 $ the above throughput, 
\vspace*{-0.1cm}
$$
T\geq K_5 (\sqrt{A_0})^{\frac{1}{2-b}-\varepsilon_1}
$$
which is the performance claimed in the lemma.

It remains to verify that we can achieve a rate $K_3M^\prime$ in the $M^\prime\times M^\prime$ MIMO transmissions between the two clusters of area $A_c$. Note that since the $M^\prime$ nodes in each group are chosen randomly among the $M$ nodes in each cluster, without any consideration on node locations, they are uniformly and independently distributed over the area $A_c$. It can be readily verified that the condition $\sqrt{A_c} \leq d \leq A_c$ in Theorem~\ref{Aoverd} is satisfied. It remains to verify that we can transmit with power $P_0/M^\prime$ such that $P_0$ satisfies (\ref{powerMIMO}). Note that due to the extra time division between the $M/M^\prime$ distinct groups in each cluster, each node is transmitting in only $M^\prime/M$ of the total transmission time of the cluster. On the other hand, due to the time sharing between the clusters in the second phase, each cluster is only active in a fraction $M/n$ of the total completion time of the phase. Therefore during the $M^\prime\times M^\prime$ MIMO transmissions, the nodes in the transmit group can transmit with elevated power $nP/M^\prime$ and still satisfy their average power constraint $P$. This, in turn, means that they can satisfy the power requirement (\ref{powerMIMO}) in Theorem~\ref{Aoverd}.

\bigbreak
\subsubsection{$n<A_0\leq n^{\frac{2(4-b)}{5-2b}}$}
In this case, we choose the cluster area as 
\begin{equation}\label{Ac_2}
A_c=A_0^{\frac{3}{4-b}}.
\end{equation} 
For this choice, the current condition $n<A_0\leq n^{\frac{2(4-b)}{5-2b}}$
on the network gives $M<A_c\leq M^2$.  This implies that, the scheme given in the hypothesis of Lemma~\ref{lem:rec} can now achieve an aggregate throughput $K_4\,\min(M,\sqrt{A_c})^{b}=(\sqrt{A_c})^b$ when used inside the clusters of area $A_c$ and number of nodes $M$. Applying exactly the scheme in the earlier case (1), we now get an aggregate throughput
$$
\frac{nM}{M^{2}A_c^{-b/2}/K_4+nM/M^\prime K_3+QM^{2}A_c^{-b/2}/K_3K_4}.
$$
The three terms in the denominator of this expression are order-wise equal for the cluster area given in \eqref{Ac_2}. Therefore, the throughput achieved is given by 
$$T =K_5M^\prime%=K_5 \frac{A_c}{\sqrt{A_0}} A_0^{-\varepsilon_2}
=K_5\,A_0^{\frac{2+b}{2(4-b)}}A_0^{-\varepsilon_1}\geq K_5 (\sqrt{A_0})^{\frac{1}{2-b}-\varepsilon_1},$$ for a constant $K_5>0$ and any $\varepsilon_1>0$. The last inequality follows from the fact that  $0\leq b < 1 $.
 %It can be verified that for $0\leq b < 1 $, 
%$$
%T\geq K_5 \frac{A_c}{\sqrt{A_0}} A_0^{-\varepsilon_2}
%$$
%which is the performance claimed in the lemma. This concludes the proof of the lemma.

Combining the conclusions of Sections~\ref{HC1} and \ref{HC2} above completes the proof of Lemma~\ref{lem:rec}.\hfill$\square$

\appendices
\section{Proof of Lemma~\ref{Aoverd}}

Lemma~\ref{Aoverd} will be proven in two steps. We first lower bound the expected capacity of the MIMO channel over random node positions and then show that for a random  realization of the node positions, the capacity of the corresponding MIMO channel is not that different from its expected value. We formally state these two results in the following lemmas.
\begin{lem}\label{Aoverd1} The expected capacity $\CM$ of the MIMO channel in Lemma~\ref{Aoverd} is lower bounded by
\begin{align*}\label{eq:thm_step1}
\EE(\CM)&=\EE(\log\det\left(I+(P_0/M)\, HH^\dagger\right))\\
&\ge K_3\min \left(M,\frac{A_c/d}{\log(A_c/d)} \right), 
\end{align*}
for a constant $K_3>0$, where the expectation is taken over the independent and uniform distribution of node positions over the transmit and receive domains of area $A_c$. 
\end{lem}
\begin{lem}\label{Aoverd2} 
Let $s=\min\left(M,\frac{A_c/d}{\log(A_c/d)}\right)$, for any $t>0$
$$
\PP\left(\left|\CM-\EE(\CM)\right|>t\right)\le e^{-\frac{2t^2}{s}}.
$$
\end{lem}

\bigbreak

Choosing $t=s^{1/2+\eps_2}, \eps_2>0$, the probability in the second lemma decreases to zero for increasing $s$. This implies that the deviations of $\CM$ from $\EE(\CM)$ are, at most, of the order of $\sqrt{s}$. Therefore combining the results of these two lemmas yields the result given in  Lemma~\ref{Aoverd}. In the sequel, we prove Lemma~\ref{Aoverd1}. The proof of Lemma~\ref{Aoverd2} closely follows the proof of Proposition~5.2 in \cite{OLP06} and is skipped due to space limitations.  

\medskip

\textit{Proof of Lemma~\ref{Aoverd1}:} For notational convenience, we start by defining 
\begin{equation}\label{fading_model2}
f_{ik}= \frac{d}{r_{ik}} \, e^{j\, 2 \pi r_{ik} }=\frac{d}{\Vert \bx_k - \bw_i \Vert}\, e^{j \,  2 \pi \Vert \bx_k - \bw_i \Vert }
\end{equation}
where $r_{ik}$ denotes the distance between the nodes $k\in\mathcal{D}_T$ and $i\in\mathcal{D}_R$ located at positions $\bx_k$ and $\bw_i$ respectively . Note that $d\le r_{ik}\le d(1+2\sqrt{2A_c}/d)$, and therefore
\begin{equation}\label{rho1}
c_0\le(1+2\sqrt{2A_c}/d)^{-1}\le |f_{ik}|\le 1, 
\end{equation}
where $c_0:=(1+2\sqrt{2})^{-1}$ and the first inequality follows from the fact that $\sqrt{A_c}\leq d$.

The first ingredient of the proof of Lemma~\ref{Aoverd1} is the Paley-Zygmund inequality used in \cite{OLT07} to prove Lemma~4.3.  We have
\begin{align*}
\EE(\CM) &= \EE \left( \log \det \left( I + \frac{P_0}{N_0W} \,\frac{1}{M} HH^\dagger \right) \right)\\
&= \EE \left( \log \det \left( I + \frac{GP_0}{N_0Wd^2} \,\frac{1}{M} FF^\dagger \right) \right)\\
& = M \, \EE\left( \log\left(1+\frac{GP_0}{N_0Wd^2}\, \lambda\right)\right) \\
&\ge M \, \log\left(1+\frac{GP_0}{N_0Wd^2}\, t\right) \, \PP (\lambda > \,t\,)
\end{align*}
for any $t>0$, where $\lambda$ is an eigenvalue of $(1/M) \,FF^\dagger$ picked uniformly at random. By Paley-Zygmund's inequality, if $0<t<\EE(\lambda)$, we have
$$
\EE(\CM) \ge M \, \log\left(1+\frac{GP_0}{N_0Wd^2}\,t\right) \, \frac{(\EE(\lambda)-t)^2}{\EE(\lambda^2)}
$$
Given \eqref{fading_model2}, we have
\begin{align*}
\EE(\lambda)&=\frac{1}{M^2} \, \EE\left(\Tr(FF^\dagger)\right) = \frac{1}{M^2} \sum_{i,k=1}^M \EE(|f_{ik}|^2)\geq c_0^2.\\
\EE(\lambda^2) &= \frac{1}{M^3} \, \EE(\Tr(FF^\dagger FF^\dagger))\\
& = \frac{1}{M^3} \sum_{i,k,l,m=1}^M \EE( f_{ik} f_{lk}^* f_{lm} f_{im}^*)\\
& \le 2+\frac{1}{M^3} \sum_{\substack{i,k,l,m=1\\i\neq l, k\neq m}}^M \EE( f_{ik} f_{lk}^* f_{lm} f_{im}^*)
\le 2 + M \, S
\end{align*}
where the last inequality follows from the upper bound in \eqref{rho1}. $S = |\EE(f_{aa} \, f_{ba}^* \, f_{bb} \, f_{ab}^*)|$ where $a$, $b$ are two different indices (notice that $S$ does not depend on the specific choice of $a$ and $b$). See Figure~\ref{four_nodes}. Choosing then $t=c_0^2/2$,
we obtain
\begin{align*}
\EE(\CM) &\ge (M \, c_0^4/4) \, \log\left(1+\frac{GP_0\,c_0^2}{2N_0Wd^2}\right) \frac{1}{2+ M \, S} \\
&\geq K_3^\prime \min \left(M, \frac{1}{S} \right)
\end{align*}
for a constant $K_3^\prime>0$ independent of $M$ and $S$ if
$$
\frac{GP_0}{N_0Wd^2}> 0\,\text{dB}.
$$ 
The quantity $S$, which takes values between $0$ and $1$, dictates therefore the capacity scaling. In the case where the channel matrix entries $f_{ik}$ are i.i.d. phases, $S=0$, so the capacity $\EE(\CM)$ is of order $M$. At the other end, if we consider the LOS channel model in \eqref{fading_model2} in the scenario where nodes are placed on a single straight line, then a simple computation shows that $
S=1$, so that $\EE(\CM)$ is of order $1$ (in this case, we know that the matrix $F$ is also rank one, so the lower bound matches the upper bound on the capacity, up to a $\log M$ term). The problem we are looking at lies between these two extremes. Our aim in the following is to show that if both $A$ and $d$ grow large and $\sqrt{A_c}\leq d \leq A_c$, then there exists $K_3^{\prime\prime}>0$ independent of $A_c$ and $d$, such that
\vspace*{-0.05cm}
\begin{equation} \label{doverA}
S \le K_3^{\prime\prime} \, \frac{d}{A_c} \, \log\left(\frac{A_c}{d}\right).
\vspace*{-0.1cm}
\end{equation}
This implies that
\vspace*{-0.2cm}
$$
\EE(\CM) \ge K_3 \, \min \left( M, \frac{A_c/d}{\log(A_c/d)} \right)
$$
which completes the proof.

\begin{figure}[tbp]
\begin{center}
\input{four_nodes.pstex_t}
\end{center}
\caption{$S = |\EE(f_{aa} \, f_{ba}^* \, f_{bb} \, f_{ab}^*)|$}
\label{four_nodes}
\end{figure}
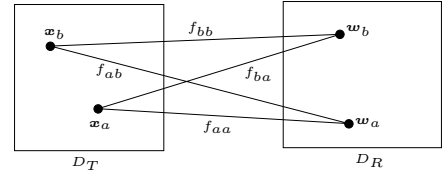
\begin{figure}[tbp]
\begin{center}
\input{coord.pstex_t}
\end{center}
\caption{Coordinate system.}
\label{coord}
\end{figure}
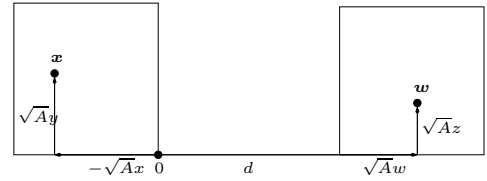

\smallbreak
The rest of the section is devoted to proving \eqref{doverA}. Let us first explicitly write the expression for $S$. We have 
\begin{align}
& S = |\EE(f_{aa} \, f_{ba}^* \, f_{bb} \, f_{ab}^*)|\nonumber\\
& = \bigg| \frac{1}{A_c^4} \int_{D_T} d\bx_a \int_{D_T} d\bx_b \int_{D_R} d\bw_a \int_{D_R} d\bw_b\,\rho\, e^{ j\,2 \pi\,\Delta}\bigg|\label{eq:S}
\end{align}
where
\vspace*{-0.1cm}
\begin{align}\label{delta}
\Delta &=
\Vert \bx_a - \bw_a \Vert - \Vert \bx_a - \bw_b \Vert
+ \Vert \bx_b - \bw_b \Vert - \Vert \bx_b - \bw_a \Vert,\\
\label{rho}
\rho&=d \left(\Vert \bx_a - \bw_a \Vert\Vert \bx_a - \bw_b \Vert\Vert \bx_b - \bw_b \Vert\Vert \bx_b - \bw_a \Vert\right)^{-1}.
\end{align}
\vspace*{-0.5cm}

We first derive the result \eqref{doverA} by approximating the distance in \eqref{fading_model2} in the regime $\sqrt{A_c}\ll d\ll A_c$. This approximate analysis captures most of the intuitions for the precise derivation which is given afterwards.  Consider two nodes at positions $\bx=(- \sqrt{A_c} \, x, \sqrt{A_c} y) \in D_T$ and $\bw=( d + \sqrt{A_c} w, \sqrt{A_c} z) \in D_R$, where $x,y,w,z \in [0,1]$ (see Figure \ref{coord}). Using the assumption that $d \gg \sqrt{A_c}$, we obtain
\begin{eqnarray*}
\Vert \bx - \bw \Vert & = & \sqrt{(d + \sqrt{A_c} \, (x+w))^2 + A_c \, (y-z)^2} \\
%& \approx & (d+\sqrt{A_c} \, (x+w)) \, \sqrt{ 1 + \frac{A_c}{d^2} \, (y-z)^2} \\ 
&\approx &d + \sqrt{A_c} \, (x+w) + \frac{A_c}{2d} \, (y-z)^2
\end{eqnarray*}
which in turn implies
\begin{align*}
& \Delta = \Vert \bx_a - \bw_a \Vert - \Vert \bx_a - \bw_b \Vert + \Vert \bx_b - \bw_b \Vert - \Vert \bx_b - \bw_a \Vert\\
& \approx \frac{A_c}{2d} \, ((y_a-z_a)^2 - (y_a-z_b)^2 + (y_b-z_b)^2 - (y_b-z_a)^2)\\
& = - \frac{A_c}{d} (y_b-y_a) \, (z_b-z_a)
\end{align*}
Next, let us also make the approximation that $\rho \approx 1$ in \eqref{rho}: this is actually assuming that the spatial degrees of freedom between the two clusters are mainly determined by the phases of the channel coefficients and not so much by the amplitudes. We will see below that this intuition is correct.

These two successive approximations lead to the following expression for $S$:
\vspace*{-0.2cm}
\begin{align*}
&S \approx S_0\\
& = \bigg| \int_0^1 dy_a \int_0^1 dy_b \int_0^1 dz_a \int_0^1 dz_b\,\, e^{ -j\, 2 \pi \frac{A_c}{d} \, (y_b-y_a) \, (z_b-z_a) } \bigg|\\
& = 2 \, \bigg| \int_0^1 dy_a \int_{y_a}^{ 1} dy_b \int_0^1 dz_a \int_0^1 dz_b \,\, e^{ - j2 \pi\frac{A_c}{d} \, (y_b-y_a) \, (z_b-z_a)}\bigg|,
\end{align*}
where the second equation follows from the symmetry of the integrand. Note that this expression does not depend on the horizontal positions of the nodes. This can be interpreted as follows. Provided the above approximation is valid, the MIMO capacity scaling between two clusters of $M$ nodes separated by a distance $d \gg \sqrt{A_c}$ is the same, be the nodes uniformly distributed on two squares of area $A_c$ or on two parallel (vertical) lines of length $\sqrt{A_c}$. This result is of interest in itself and can be proven rigorously. %in Appendix \ref{aux} (Proposition \ref{parallel_lines}).

We show below that the above integral is indeed of order $d/A_c$. Let us compute the first integral, which yields
\begin{align*}
\int_0^1 dz_b &\, e^{ - 2 \pi j \, \frac{A_c}{d} \, (y_b-y_a) \, (z_b-z_a)}\\
&= - \frac{d}{j2 \pi A_c \, (y_b-y_a)} \, e^{ -j\, 2 \pi \frac{A_c}{d} \, (y_b-y_a) \, (z_b-z_a)} \bigg|_{z_b=0}^{z_b=1}.
\end{align*}
This implies that
\begin{align*}
\left| \int_0^1 dz_b \,e^{ -j\, 2 \pi \frac{A_c}{d} \, (y_b-y_a) \, (z_b-z_a)} \right|\le K_6 \, \frac{d}{A_c} \, \frac{1}{|y_b-y_a|}
\end{align*}
for a constant $K_6$ independent of $A_c$ and $d$. We can divide the integration over $y_a$ and $y_b$ into two parts,
\begin{align*}
&\int_0^1 dy_a \int_{y_a}^{ 1} dy_b \int_0^1 dz_a \int_0^1 dz_b \,\, e^{ - j2 \pi\frac{A_c}{d} \, (y_b-y_a) \, (z_b-z_a)} \\
&=\left(\int_0^1 dy_a \int_{y_a}^{(y_a+\eps_3) \vee 1} dy_b\,\,+\,\,\int_0^{1-\eps_3} dy_a \int_{y_a+\eps_3}^1\right)\\
&\hspace*{2cm}\times \int_0^1 dz_a \int_0^1 dz_b \,\,e^{ - j2 \pi\frac{A_c}{d} \, (y_b-y_a) \, (z_b-z_a)}, 
\end{align*} 
for any $0<\eps_3<1$. The first term can be simply bounded by $\eps_3$, which yields the following upper bound for $S_0$ 
\begin{align*}
S_0 &\le  2 \eps_3 + 2K_6 \, \frac{d}{A_c} \int_0^{1-\eps_3} dy_a \int_{y_a+\eps_3}^1 dy_b \, \frac{1}{|y_b-y_a|}\\
& \le 2 \varepsilon + 2K_6 \, \frac{d}{A_c} \, \log(1/\eps_3)
\end{align*}
So choosing $\eps_3=d/A_c$, we finally obtain
$$
S \approx S_0 \le K_3^{\prime\prime} \, \frac{d}{A_c} \, \log(A_c/d)
$$
for a constant $K_3^{\prime\prime}$ independent of $A_c$ and $d$. We will next prove \eqref{doverA}  without making use of the above approximations.

%%%%%%%%%%%%%%%%%%%%%%%%%%%%%%%%%%%%%%%%%%%
%%%%%%%%%%%%%%%%%%%%%%%%%%%%%%%%%%%%%%%%%%%
%%%%%%%%%%%%%%%%%%%%%%%%%%%%%%%%%%%%%%%%%%%

{\it Proof of Inequality~\eqref{doverA}:}
We start again with the expression for $S$ in \eqref{eq:S}. Note that due to the symmetry of $\Delta$ and $\rho$ in $\bw_a $ and $\bw_b$, we can upper bound \eqref{eq:S} as 
\begin{eqnarray*}
S\le \frac{d^4}{A_c^4} \int_{D_T} d\bx_a \int_{D_T} d\bx_b \left| \int_{D_R} d\bw \,
\frac{\,\,e^{ j\,2 \pi \left(\Vert \bx_a - \bw \Vert - \Vert \bx_b - \bw \Vert\right) }}{\Vert \bx_a - \bw \Vert\,\Vert \bx_b - \bw \Vert} \right|^2
\end{eqnarray*}
Expressing this upper bound more explicitly in the coordinate system in Figure~\ref{coord}, we obtain the following upper bound for $S$,
\begin{equation}
\int_0^1 dx_a \int_0^1 dy_a \int_0^1 dx_b \int_0^1 dy_b  \left|
\int_0^1 dw \int_0^1 dz \,\frac{e^{ j\,2 \pi \, g_{a,b}(w,z)}}{G_{a,b}(w,z)} \right|^2\label{ub2}
\end{equation}
where
\begin{align*}
g_{a,b}(w,z) =&\sqrt{(d+\sqrt{A_c} \, (x_a+w))^2 + A_c \, (y_a-z)^2}\\
&- \sqrt{(d+\sqrt{A_c} \, (x_b+w))^2 + A_c \, (y_b-z)^2}.
\end{align*}
and
\begin{align*}
G_{a,b}(w,z)& =d^{-2}\sqrt{(d+\sqrt{A_c} \, (x_a+w))^2 + A_c \, (y_a-z)^2}\\ &\quad \times \sqrt{(d+\sqrt{A_c} \, (x_b+w))^2 + A_c \, (y_b-z)^2}.
\end{align*}

Let us first focus on the integral inside the square in \eqref{ub2}. The key idea behind the next steps of the proof is contained in the following two lemmas.

\begin{lem} \label{mu}
Let $g:[0,1] \to \RR$ be a $C^2$ function such that $|g'(z)| \ge c_1>0$ for all $z \in [0,1]$ and $g''$ changes sign at most twice on $[0,1]$ (say e.g.~$g''(z) \ge 0 $ in $[z_-,z_+]$ and $g''(z) \le 0$ outside). Let also $G:[0,1] \to \RR$ be a $C^1$ function such that $|G(z)| \ge c_2>0$ and $G'(z)$ changes sign at most twice on $[0,1]$. Then %there exists $K_{11}>0$ independent of $\mu$ and $c_1$ such that
$$
\left| \int_0^1 dz \, \frac{e^{ j\,2 \pi g(z)}}{G(z)} \right| \le \frac{14}{\pi \, c_1 \, c_2}.% \quad \forall \mu>0
$$
\end{lem}

\medbreak

\begin{lem} \label{sqrt_mu}
Let $g:[0,1] \to \RR$ be a $C^2$ function such that there exists $z_0 \in [0,1]$ and $c_1>0$ with $|g'(z)| \ge c_1 \, |z-z_0|$ for all $z \in [0,1]$ and $g''$ changes sign at most twice on $[0,1]$. Let also $G:[0,1] \to \RR$ be a $C^1$ function such that $|G(z)| \ge c_2>0$ and $G'(z)$ changes sign at most twice on $[0,1]$. Then %there exists $K_{12}>0$ independent of $\mu$ and $c_1$ such that
$$
\left| \int_0^1 dz \, \frac{e^{ j\,2 \pi\, g(z)}}{G(z)}  \right| \le \sqrt{\frac{14}{\pi\, c_1 \, c_2}}. % \quad \forall \mu>0
$$
\end{lem}

The proof of Lemma \ref{mu} is relegated to Appendix~\ref{tech}. The proof of Lemma \ref{sqrt_mu} follows the same lines and is omitted due to space limitations.

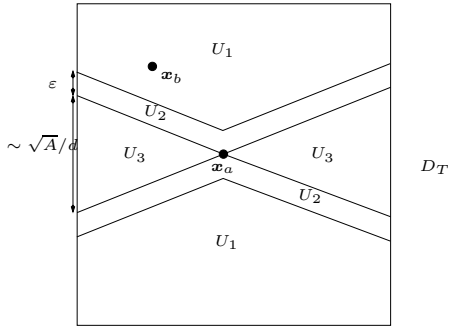
\begin{figure}[hbt]
\begin{center}
\input{regions.pstex_t}
\end{center}
\caption{Domains of integration: the relative positions of the points $\bx_a$ and $\bx_b$ determine in which domain one is ($U_1$ on the figure).}
\label{regions}
\end{figure}

Let now $\eps_3>0$ and let us divide the integration domain $(x_a,x_b,y_a,y_b) \in [0,1]^4$  in \eqref{ub2} into three subdomains (see Figure \ref{regions}):
\begin{eqnarray*}
U_1 & = & \left\{  |y_a-y_b| - (\sqrt{A_c}/d)\, |x_b-x_a| \ge \eps_3 \right\}\\
U_2 & = & \left\{ 0 < |y_a-y_b| -  (\sqrt{A_c}/d)\, |x_b-x_a| < \eps_3 \right\}\\
U_3 & = & \left\{ |y_a-y_b| \le (\sqrt{A_c}/d)\, |x_b-x_a| \right\}
\end{eqnarray*}
Consider first the integral over $U_1$. It can be verified from the expression \eqref{g'} for the first order partial derivative of $g_{a,b}$ with respect to $z$ given in Appendix~\ref{tech} that if $(x_a,x_b,y_a,y_b) \in U_1$, then
$$
\left| \frac{\partial g_{a,b}}{\partial z} (w,z) \right|
\ge K_7 \,\frac{A_c}{d} \left( |y_b-y_a| - \frac{\sqrt{A_c}}{d} \, |x_b-x_a| \right)
$$
for a constant $K_7>0$ independent of $A_c$ and $d$. Notice next that $|G_{a,b}(y,z)| \ge 1$. It can further be checked that both $\frac{\partial^2 g_{a,b}}{\partial z^2} (w,z)$ and $\frac{\partial G_{a,b}}{\partial z}(w,z)$ change sign at most twice on the interval $z \in [0,1]$ (for $w$ fixed). Therefore,
applying Lemma \ref{mu}, we conclude that
\begin{align*}
& \left| \int_0^1 dw \int_0^1 dz \,\frac{e^{ j\,2 \pi \, g_{a,b}(w,z)}}{G_{a,b}(z)} \right|
\le \int_0^1 dw \left| \int_0^1 dz \, \frac{e^{ j\,2 \pi \, g_{a,b}(w,z)}}{G_{a,b}(y,z)} \right|\\
&\hspace*{2cm} \le K_8 \, \frac{d}{A_c} \, \frac{1}{|y_b-y_a| - (\sqrt{A_c}/d) \, |x_b-x_a|}
\end{align*}
Since we know that this integral is also less than $1$, this in turn implies
\begin{align*}
&\int_{U_1} dx_a dx_b dy_a dy_b \, \left| \int_0^1 dw \int_0^1 dz \, \frac{e^{j\,2 \pi \; g_{a,b}(w,z)}}{G_{a,b}(w,z)} \right|^2\\
& \le K_8 \, \frac{d}{A_c} \int_{U_1} dx_a dx_b dy_a dy_b \, \frac{1}{|y_b-y_a| - (\sqrt{A_c}/d) \, |x_b-x_a|} \\
& = K_8 \, \frac{d}{A_c} \, \log (1/\eps_3)
\end{align*}

Second, it is easy to check that
\begin{align*}
\int_{U_2} dx_a dx_b dy_a dy_b\left| \int_0^1 dw \int_0^1 dz \, \frac{e^{j \, 2 \pi  \,  g_{a,b}(w,z)}}{G_{a,b}(w,z)} \right|^2 \le 2 \eps_3.
\end{align*}

The integral over the third domain of integration $U_3$ is more delicate. Notice first that the obvious bound
$$
\int_{U_3} dx_a dx_b dy_a dy_b \, \left| \int_0^1 dw \int_0^1 dz \, \frac{e^{ j \;2 \pi \, g_{a,b}(w,z)}}{G_{a,b}(w,z)} \right|^2 \le 2 \, \frac{\sqrt{A_c}}{d}
$$
allows to obtain
$$
S \le K_8 \, \frac{d}{A_c} \, \log (1/\eps_3) + 2\eps_3 + 2 \, \frac{\sqrt{A_c}}{d}
$$
which can be made smaller than $K_3 \, (d/A_c) \, \log(A_c/d)$ by choosing $\eps_3=d/A_c$ when $A_c^{3/4} \le d \le A_c$ (as $\sqrt{A_c}/d \le d/A_c$ in this case).

For the remainder of the proof, let us therefore assume that $\sqrt{A_c} \le d \le A_c^{3/4}$. As before, we focus on the integral inside the square in the following term
\begin{equation} \label{ub_h}
\int_{U_3} dx_a dx_b dy_a dy_b \, \left| \int_0^1 dw \int_0^1 dz \, \frac{e^{ j 2 \pi\; g_{a,b}(w,z)}}{G_{a,b}(w,z)} \right|^2.
\end{equation}
Let us start by considering the simplest case where the points $\bx_a$ and $\bx_b$ are located on the same horizontal line, i.e.~$y_a=y_b$. In this case, the second term in the expression \eqref{g'} for $\frac{\partial g_{a,b}}{\partial z} (w,z)$ becomes zero, so we deduce the following lower bound:
$$
\left| \frac{\partial g_{a,b}}{\partial z} (w,z) \right| \ge K_{9} \,\frac{A_c^{3/2}}{d^2} |x_b-x_a| \, |z-y_a|
$$
This, together with the above mentioned properties of the functions $g_{a,b}$ and $G_{a,b}$, allows us to apply Lemma \ref{sqrt_mu} so as to obtain
$$
\left| \int_0^1 dw \int_0^1 dz \, \frac{e^{ j 2 \pi\, g_{a,b}(w,z)}}{G_{a,b}(w,z)} \right| 
\le K_{10} \, \frac{d}{A_c^{3/4}} \, \frac{1}{\sqrt{|x_b-x_a|}}
$$
for a constant $K_{10}>0$ independent of $A_c$ and $d$.
A slight generalization of this argument (see Appendix~\ref{tech} for details) shows that not only when $y_a=y_b$ but for any $(x_a,x_b,y_a,y_b) \in U_3$, we have
\begin{align}
&\left| \int_0^1 dw \int_0^1 dz \, \frac{e^{ j 2 \pi\, g_{a,b}(w,z)}}{G_{a,b}(w,z)} \right|\nonumber\\ 
&\hspace*{2cm} \le K_{10} \, \frac{d}{A_c^{3/4}} \, \frac{1}{((x_b-x_a)^2+(y_b-y_a)^2)^{1/4}} \nonumber\\
&\hspace*{2cm} \le K_{10} \, \frac{d}{A_c^{3/4}} \, \frac{1}{\sqrt{|x_b-x_a|}} \label{lb}
\end{align}
Since we also know that the above integral is less than $1$, we further obtain
\begin{align*}
\left| \int_0^1 dw \int_0^1 dz \, \frac{e^{ j 2 \pi\, g_{a,b}(w,z)}}{G_{a,b}(w,z)} \right|^2\le \min \left( K_{10} \, \frac{d^2}{A_c^{3/2}} \, \frac{1}{|x_b-x_a|} \; , \; 1 \right)
\end{align*}
For any $0<\eta<1$, we can now upper bound \eqref{ub_h} as
\begin{align*}
&\int_{U_3} dx_a dx_b dy_a dy_b \left| \int_0^1 dw \int_0^1 dz \, \frac{e^{ j 2 \pi\, g_{a,b}(w,z)}}{G_{a,b}(w,z)} \right|^2\\
& \le|U_3 \cap \{|x_b-x_a| < \eta\}|\\
& + K_{10} \int_{U_3 \cap \{|x_b-x_a| \ge \eta\}} dx_a dx_b dy_a dy_b \, \frac{d^2}{A_c^{3/2}} \, \frac{1}{|x_b-x_a|}\\
& \le 2 \eta + K_{10} \, \frac{\sqrt{A_c}}{d} \, \frac{d^2}{A_c^{3/2}} \, \log(1/\eta) = 2 \eta + K_{10}\, \frac{A_c}{d} \, \log(1/\eta)
\end{align*}
implying that
$$
S \le K_8 \, \frac{d}{A_c} \, \log (1/\eps_3) + 2\eps_3 + 2\eta + K_{10} \, \frac{d}{A_c} \, \log (1/\eta)
$$
Choosing finally $\eps_3=\eta=d/A_c$ allows to conclude that $S \le K \, (d/A_c) \, \log(A_c/d)$ also in the case where $\sqrt{A_c} \le d \le A_c^{3/4}$. \hfill $\square$\\

%%%%%%%%%%%%%%%%%%%%%%%%%%%%%%%%%%%%%%%%%%%%
%%%%%%%%%%%%%%%%%%%%%%%%%%%%%%%%%%%%%%%%%%%%
%%%%%%%%%%%%%%%%%%%%%%%%%%%%%%%%%%%%%%%%%%%%

\section{Technical details} \label{tech}

{\it Proof of Lemma \ref{mu}.}
By the integration by parts formula, we obtain
\begin{align*}
&\int_0^1 dz \, \frac{e^{j2 \pi g(z)}}{G(z)} =
\int_0^1 dz \, \frac{j\,2 \pi g'(z)}{j\,2 \pi  g'(z)G(z)} \, e^{ 2 \pi j  g(z)}\\
&\hspace*{-0.2cm} =  \frac{e^{ j\,2 \pi g(z))}}{j\,2 \pi  g'(z)G(z)} \bigg|_0^1 - \int_0^1 dz \, \frac{g''(z)G(z)+g'(z)G'(z)}{j\,2 \pi (g'(z)G(z))^2} \, e^{ j\,2 \pi  g(z)}
\end{align*}
which in turn yields the upper bound
\begin{align*}
&\left|\int_0^1 dz \, \frac{e^{j2 \pi g(z)}}{G(z)} \right|\le \frac{1}{2 \pi } \, \Bigg( \frac{1}{|g'(1)||G(1)|} + \frac{1}{|g'(0)||G(0)|}\\
& + \int_0^1 dz \, \frac{|g''(z)|}{(g'(z))^2|G(z)|} + \int_0^1 dz \, \frac{|G'(z)|}{g'(z)(G(z))^2} \Bigg).
\end{align*}
By the assumptions made in the lemma, we have
\begin{align*}
&\int_0^1 dz \, \frac{|g''(z)|}{(g'(z))^2|G(z)|} \le \frac{1}{c_2} \int_0^1 dz \, \frac{|g''(z)|}{(g'(z))^2}\\
&\hspace*{1cm} = \frac{1}{c_2} \Bigg( - \int_0^{z_-} dz \, \frac{g''(z)}{(g'(z))^2} + \int_{z_-}^{z_+} dz \, \frac{g''(z)}{(g'(z))^2}\\
&\hspace*{2cm} - \int_{z_+}^1 dz \, \frac{g''(z)}{(g'(z))^2} \Bigg)\\
&\hspace*{1cm} = \frac{1}{c_2} \, \Bigg(
\frac{1}{g'(1)} - \frac{1}{g'(0)} + \frac{2}{g'(z_-)} - \frac{2}{g'(z_+)} \Bigg).
\end{align*}
So
$$
\int_0^1 dz \, \frac{|g''(z)|}{(g'(z))^2|G(z)|} \le \frac{6}{c_1 \, c_2}.
$$
We obtain in a similar manner that
$$
\int_0^1 dz \, \frac{|G'(z)|}{g'(z)(G(z))^2} \le \frac{6}{c_1 \, c_2}
$$
Combining all the bounds, we finally get
$$
\left|\int_0^1 dz \, \frac{e^{j2 \pi g(z)}}{G(z)} \right|  \le \frac{14}{\pi \, c_1 \, c_2}.
$$
\hfill $\square$\\

{\it Expression for the first order derivative of $g_{a,b}(w,z)$:} It can be verified that
\begin{align}
g_{a,b}&(w,z)=-\sqrt{A_c}\int_{x_a}^{x_b} \frac{(d/\sqrt{A_c}+x+w)\,\, dx}{\sqrt{(d/\sqrt{A_c}+x+w)^2 + (y_a-z)^2}}\nonumber\\
&  + \sqrt{A_c}\int_{y_a}^{y_b} \frac{(y-z)\,\,dy}{\sqrt{(d/\sqrt{A_c}+x_b+w)^2 + (y-z)^2}} \label{g}
\end{align}
So the expression for the first order partial derivative of $g_{a,b}(w,z)$ with respect to $z$ is given by
\begin{align}
&\hspace*{-0.3cm} \frac{\partial g_{a,b}}{\partial z} (w,z)
= \sqrt{A_c} \int_{x_a}^{x_b} \frac{(z-y_a) \, (d/\sqrt{A_c}+x+w)\,\,\, dx}{\left((d/\sqrt{A_c}+x+w)^2+(z-y_a)^2\right)^{3/2}}\nonumber\\
&\hspace*{0.5cm}   + \sqrt{A_c}\int_{y_a}^{y_b}\frac{(d/\sqrt{A_c}+x_b+w)^2\,\,\,  dy }{\left((d/\sqrt{A_c}+x_b+w)^2+(z-y)^2\right)^{3/2}}\label{g'}
\end{align}

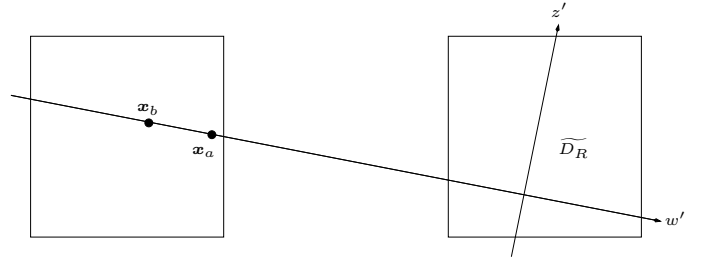
\begin{figure}[tbp]
\begin{center}
\input{tilt.pstex_t}
\end{center}
\caption{Tilted reference frame.}
\label{tilt}
\end{figure}

{\it Proof of equation \eqref{lb}:} In order to prove \eqref{lb}, we need to make a change of coordinate system, replacing $(w,z)$ by $(w',z')$, where $w'$ is now in the direction of the vector $\bx_a-\bx_b$ and $z'$ is perpendicular to it (see Figure~\ref{tilt} ). In this new coordinate system, the integral reads
$$
\left| \int_{\widetilde{D_R}} dw' dz' \, \frac{e^{j\, 2 \pi g_{a,b}(w',z')}}{G_{a,b}(w',z')} \right| 
$$
where $g_{a,b}(w',z')$, $G_{a,b}(w',z')$ have the same form as $g_{a,b}(w,z)$, $G_{a,b}(w,z)$, but now, the domain of integration $\widetilde{D_R}$ is a tilted square, as indicated on the Figure~\ref{tilt}. Using then the same argument as in the case $y_a=y_b$, we conclude that
$$
\left| \int_{\widetilde{D_R}} dw' dz' \, \frac{e^{j\, 2 \pi g_{a,b}(w',z')}}{G_{a,b}(w',z')} \right| 
\le K_{10} \, \frac{d}{A_c^{3/4}} \, \frac{1}{\sqrt{|x_b'-x_a'|}}.
$$
Noticing finally that $|x_b'-x_a'|=\sqrt{(x_b-x_a)^2+(y_b-y_a)^2}$ allows to conclude \eqref{lb}.

\end{document}

%% file: squares.pstex_t
\begin{picture}(0,0)%
\includegraphics{squares.pstex}%
\end{picture}%
\setlength{\unitlength}{2487sp}%
\begingroup\makeatletter\ifx\SetFigFont\undefined%
\gdef\SetFigFont#1#2#3#4#5{%
  \reset@font\fontsize{#1}{#2pt}%
  \fontfamily{#3}\fontseries{#4}\fontshape{#5}%
  \selectfont}%
\fi\endgroup%
\begin{picture}(4902,2199)(1936,-2473)
\put(2476,-2386){\makebox(0,0)[lb]{\smash{{\SetFigFont{9}{10.8}{\familydefault}{\mddefault}{\updefault}{\color[rgb]{0,0,0}$\sqrt{A}$}%
}}}}
\put(5776,-511){\makebox(0,0)[lb]{\smash{{\SetFigFont{9}{10.8}{\familydefault}{\mddefault}{\updefault}{\color[rgb]{0,0,0}$D_R$}%
}}}}
\put(4276,-1036){\makebox(0,0)[lb]{\smash{{\SetFigFont{9}{10.8}{\familydefault}{\mddefault}{\updefault}{\color[rgb]{0,0,0}$d$}%
}}}}
\put(1951,-886){\makebox(0,0)[lb]{\smash{{\SetFigFont{9}{10.8}{\familydefault}{\mddefault}{\updefault}{\color[rgb]{0,0,0}$D_T$}%
}}}}
\put(5701,-2386){\makebox(0,0)[lb]{\smash{{\SetFigFont{9}{10.8}{\familydefault}{\mddefault}{\updefault}{\color[rgb]{0,0,0}$\sqrt{A}$}%
}}}}
\end{picture}%

%% file: four_nodes.pstex_t
\begin{picture}(0,0)%
\includegraphics{four_nodes.pstex}%
\end{picture}%
\setlength{\unitlength}{987sp}%
\begingroup\makeatletter\ifx\SetFigFont\undefined%
\gdef\SetFigFont#1#2#3#4#5{%
  \reset@font\fontsize{#1}{#2pt}%
  \fontfamily{#3}\fontseries{#4}\fontshape{#5}%
  \selectfont}%
\fi\endgroup%
\begin{picture}(10749,4233)(1489,-6832)
\put(7276,-4486){\makebox(0,0)[lb]{\smash{{\SetFigFont{5}{6.0}{\familydefault}{\mddefault}{\updefault}{\color[rgb]{0,0,0}$f_{ba}$}%
}}}}
\put(5851,-3361){\makebox(0,0)[lb]{\smash{{\SetFigFont{5}{6.0}{\familydefault}{\mddefault}{\updefault}{\color[rgb]{0,0,0}$f_{bb}$}%
}}}}
\put(10051,-5686){\makebox(0,0)[lb]{\smash{{\SetFigFont{5}{6.0}{\familydefault}{\mddefault}{\updefault}{\color[rgb]{0,0,0}$\bw_a$}%
}}}}
\put(2926,-6736){\makebox(0,0)[lb]{\smash{{\SetFigFont{5}{6.0}{\familydefault}{\mddefault}{\updefault}{\color[rgb]{0,0,0}$D_T$}%
}}}}
\put(10051,-6661){\makebox(0,0)[lb]{\smash{{\SetFigFont{5}{6.0}{\familydefault}{\mddefault}{\updefault}{\color[rgb]{0,0,0}$D_R$}%
}}}}
\put(9826,-3361){\makebox(0,0)[lb]{\smash{{\SetFigFont{5}{6.0}{\familydefault}{\mddefault}{\updefault}{\color[rgb]{0,0,0}$\bw_b$}%
}}}}
\put(2251,-3436){\makebox(0,0)[lb]{\smash{{\SetFigFont{5}{6.0}{\familydefault}{\mddefault}{\updefault}{\color[rgb]{0,0,0}$\bx_b$}%
}}}}
\put(3376,-5761){\makebox(0,0)[lb]{\smash{{\SetFigFont{5}{6.0}{\familydefault}{\mddefault}{\updefault}{\color[rgb]{0,0,0}$\bx_a$}%
}}}}
\put(6226,-5836){\makebox(0,0)[lb]{\smash{{\SetFigFont{5}{6.0}{\familydefault}{\mddefault}{\updefault}{\color[rgb]{0,0,0}$f_{aa}$}%
}}}}
\put(3526,-4411){\makebox(0,0)[lb]{\smash{{\SetFigFont{5}{6.0}{\familydefault}{\mddefault}{\updefault}{\color[rgb]{0,0,0}$f_{ab}$}%
}}}}
\end{picture}%

%% file: coord.pstex_t
\begin{picture}(0,0)%
\includegraphics{coord.pstex}%
\end{picture}%
\setlength{\unitlength}{1224sp}%
\begingroup\makeatletter\ifx\SetFigFont\undefined%
\gdef\SetFigFont#1#2#3#4#5{%
  \reset@font\fontsize{#1}{#2pt}%
  \fontfamily{#3}\fontseries{#4}\fontshape{#5}%
  \selectfont}%
\fi\endgroup%
\begin{picture}(9549,3562)(1639,-7211)
\put(4501,-7111){\makebox(0,0)[lb]{\smash{{\SetFigFont{6}{7.2}{\familydefault}{\mddefault}{\updefault}{\color[rgb]{0,0,0}$0$}%
}}}}
\put(6301,-7111){\makebox(0,0)[lb]{\smash{{\SetFigFont{6}{7.2}{\familydefault}{\mddefault}{\updefault}{\color[rgb]{0,0,0}$d$}%
}}}}
\put(9901,-6286){\makebox(0,0)[lb]{\smash{{\SetFigFont{6}{7.2}{\familydefault}{\mddefault}{\updefault}{\color[rgb]{0,0,0}$\sqrt{A} z$}%
}}}}
\put(8701,-7111){\makebox(0,0)[lb]{\smash{{\SetFigFont{6}{7.2}{\familydefault}{\mddefault}{\updefault}{\color[rgb]{0,0,0}$\sqrt{A} w$}%
}}}}
\put(1726,-6061){\makebox(0,0)[lb]{\smash{{\SetFigFont{6}{7.2}{\familydefault}{\mddefault}{\updefault}{\color[rgb]{0,0,0}$\sqrt{A} y$}%
}}}}
\put(3151,-7111){\makebox(0,0)[lb]{\smash{{\SetFigFont{6}{7.2}{\familydefault}{\mddefault}{\updefault}{\color[rgb]{0,0,0}$-\sqrt{A} x$}%
}}}}
\put(2401,-4861){\makebox(0,0)[lb]{\smash{{\SetFigFont{6}{7.2}{\familydefault}{\mddefault}{\updefault}{\color[rgb]{0,0,0}$\bx$}%
}}}}
\put(9751,-5461){\makebox(0,0)[lb]{\smash{{\SetFigFont{6}{7.2}{\familydefault}{\mddefault}{\updefault}{\color[rgb]{0,0,0}$\bw$}%
}}}}
\end{picture}%

%% file: regions.pstex_t
\begin{picture}(0,0)%
\includegraphics{regions.pstex}%
\end{picture}%
\setlength{\unitlength}{1381sp}%
\begingroup\makeatletter\ifx\SetFigFont\undefined%
\gdef\SetFigFont#1#2#3#4#5{%
  \reset@font\fontsize{#1}{#2pt}%
  \fontfamily{#3}\fontseries{#4}\fontshape{#5}%
  \selectfont}%
\fi\endgroup%
\begin{picture}(7455,5799)(1261,-7573)
\put(4951,-4786){\makebox(0,0)[lb]{\smash{{\SetFigFont{6}{7.2}{\familydefault}{\mddefault}{\updefault}{\color[rgb]{0,0,0}$\bx_a$}%
}}}}
\put(3751,-3811){\makebox(0,0)[lb]{\smash{{\SetFigFont{6}{7.2}{\familydefault}{\mddefault}{\updefault}{\color[rgb]{0,0,0}$U_2$}%
}}}}
\put(6526,-5311){\makebox(0,0)[lb]{\smash{{\SetFigFont{6}{7.2}{\familydefault}{\mddefault}{\updefault}{\color[rgb]{0,0,0}$U_2$}%
}}}}
\put(6751,-4561){\makebox(0,0)[lb]{\smash{{\SetFigFont{6}{7.2}{\familydefault}{\mddefault}{\updefault}{\color[rgb]{0,0,0}$U_3$}%
}}}}
\put(3376,-4561){\makebox(0,0)[lb]{\smash{{\SetFigFont{6}{7.2}{\familydefault}{\mddefault}{\updefault}{\color[rgb]{0,0,0}$U_3$}%
}}}}
\put(5026,-6136){\makebox(0,0)[lb]{\smash{{\SetFigFont{6}{7.2}{\familydefault}{\mddefault}{\updefault}{\color[rgb]{0,0,0}$U_1$}%
}}}}
\put(4951,-2686){\makebox(0,0)[lb]{\smash{{\SetFigFont{6}{7.2}{\familydefault}{\mddefault}{\updefault}{\color[rgb]{0,0,0}$U_1$}%
}}}}
\put(2026,-3286){\makebox(0,0)[lb]{\smash{{\SetFigFont{6}{7.2}{\familydefault}{\mddefault}{\updefault}{\color[rgb]{0,0,0}$\eps$}%
}}}}
\put(4051,-3136){\makebox(0,0)[lb]{\smash{{\SetFigFont{6}{7.2}{\familydefault}{\mddefault}{\updefault}{\color[rgb]{0,0,0}$\bx_b$}%
}}}}
\put(8701,-4786){\makebox(0,0)[lb]{\smash{{\SetFigFont{6}{7.2}{\familydefault}{\mddefault}{\updefault}{\color[rgb]{0,0,0}$D_T$}%
}}}}
\put(1276,-4411){\makebox(0,0)[lb]{\smash{{\SetFigFont{6}{7.2}{\familydefault}{\mddefault}{\updefault}{\color[rgb]{0,0,0}$\sim\sqrt{A}/d$}%
}}}}
\end{picture}%

%% file: tilt.pstex_t
\begin{picture}(0,0)%
\includegraphics{tilt.pstex}%
\end{picture}%
\setlength{\unitlength}{1302sp}%
\begingroup\makeatletter\ifx\SetFigFont\undefined%
\gdef\SetFigFont#1#2#3#4#5{%
  \reset@font\fontsize{#1}{#2pt}%
  \fontfamily{#3}\fontseries{#4}\fontshape{#5}%
  \selectfont}%
\fi\endgroup%
\begin{picture}(12477,4806)(439,-7048)
\put(10726,-2461){\makebox(0,0)[lb]{\smash{{\SetFigFont{6}{7.2}{\familydefault}{\mddefault}{\updefault}{\color[rgb]{0,0,0}$z'$}%
}}}}
\put(12901,-6436){\makebox(0,0)[lb]{\smash{{\SetFigFont{6}{7.2}{\familydefault}{\mddefault}{\updefault}{\color[rgb]{0,0,0}$w'$}%
}}}}
\put(3901,-5086){\makebox(0,0)[lb]{\smash{{\SetFigFont{6}{7.2}{\familydefault}{\mddefault}{\updefault}{\color[rgb]{0,0,0}$\bx_a$}%
}}}}
\put(2851,-4261){\makebox(0,0)[lb]{\smash{{\SetFigFont{6}{7.2}{\familydefault}{\mddefault}{\updefault}{\color[rgb]{0,0,0}$\bx_b$}%
}}}}
\put(10876,-5086){\makebox(0,0)[lb]{\smash{{\SetFigFont{6}{7.2}{\familydefault}{\mddefault}{\updefault}{\color[rgb]{0,0,0}$\widetilde{D_R}$}%
}}}}
\end{picture}%